\documentclass{aa}  
\usepackage{graphicx}
\usepackage{txfonts}
\usepackage[utf8]{inputenc}
\usepackage{amssymb,amsmath}
\usepackage{lscape}
\usepackage{longtable}
\usepackage{float}
\usepackage{natbib}
\usepackage{color}
\begin{document} 

%
%
\title{
Gas fractions and depletion times in galaxies with different degrees of interaction
}
  \author{S. Díaz-García\inst{1,2}
          \and        
          J. H. Knapen\inst{1,2}
          }
  \institute{Instituto de Astrof\'isica de Canarias, E-38205, La Laguna, Tenerife, Spain \\
              \email{simondiazgar@gmail.com}
         \and
             Departamento de Astrof\'isica, Universidad de La Laguna, E-38205, La Laguna, Tenerife, Spain
          }
  \date{Received 20 Dec 2019 / Accepted 21 Feb 2020}
  \abstract
{
\emph{Context}. 
A moderate enhancement of the star formation rates (SFR) in local interacting galaxies has been reported, 
but the physical mechanisms leading to this increase are not clear.\\
\emph{Aims}. We study the atomic gas content and the central stellar mass concentration for a sample of almost 1500 nearby galaxies 
to further investigate the nature of starbursts and the influence of galaxy-galaxy interactions on star formation.\\
\emph{Methods}. We used a sample of catalogued interacting and non-interacting galaxies in the S$^4$G survey---along 
with archival H{\sc\,i} gas masses, stellar masses ($M_{\ast}$), 
and SFRs from \emph{IRAS} far-infrared fluxes---and calculate depletion times ($\tau$) and gas fractions. 
We traced the central stellar mass concentration from the inner slope of 
the stellar component of the rotation curves, $d_{\rm R}v_{\ast}(0)$. 
Starbursts are defined as galaxies with a factor $>4$ enhanced SFR 
relative to a control sample of non-interacting galaxies which are $\pm 0.2$ dex in stellar mass and $\pm 1$ in $T$-type.\\ 
\emph{Results}. Starbursts are mainly early-type ($T\lesssim 5$), 
massive spiral galaxies ($M_{\ast}\gtrsim 10^{10}M_{\odot}$) that are not necessarily interacting. 
For a given stellar mass bin, starbursts are characterised by lower gas depletion times, 
similar gas fractions, and larger central stellar mass concentrations than non-starburst galaxies. 
The global distributions of gas fraction and gas depletion time of interacting galaxies are not 
statistically different from those of their non-interacting counterparts. 
However, in the case of currently merging galaxies, the median gas depletion time is a factor of $0.4 \pm 0.2$ 
that of control sample galaxies, and their SFRs are a factor of $1.9 \pm 0.5$ enhanced, 
even though the median gas fraction is similar.\\
\emph{Conclusions}. 
Starbursts present long-lasting star formation 
in circumnuclear regions, which causes an enhancement of the central stellar density at $z\approx0$ in both interacting and 
non-interacting systems. Starbursts have low gas depletion timescales, yet similar gas fractions as normal main-sequence galaxies. 
Galaxy mergers cause a moderate enhancement of the star formation efficiency.
}
\maketitle
%
%
%
\section{Introduction}\label{intro}
%
%
There is plenty of observational work in the literature reporting a moderate statistical increase (by factor of a few) of the star formation rate (SFR) 
in interacting galaxies 
\citep[e.g.][]{1978ApJ...219...46L,2003A&A...405...31B,2007AJ....133..791S,2007AJ....134..527W,2008MNRAS.385.1903L,2009ApJ...704..324R,2009ApJ...698.1437K,2013MNRAS.435.3627E,2015A&A...579A..45B,2015ApJS..218....6B,2015MNRAS.454.1742K}. 
It is known that the most extreme starbursts, such as ultraluminous infrared galaxies (ULIRGs), are almost always 
interacting or merging \citep[e.g.][]{1985MNRAS.214...87J}; it is likely that this interaction stimulates the high SFR in such rare objects. 
The SFR enhancement has been found to be larger for smaller nuclear separation in galaxy pairs \citep[][]{2019ApJ...881..119P}. 
However, \citet[][]{2019A&A...631A..51P} conclude that the SFR 
of merging galaxies is not significantly different from the SFR of non-merging galaxies 
based on the use of convolutional neural networks applied to over 200000 galaxies. 
The astrophysics behind the SFR enhancement in interacting galaxies remains a matter of intense investigation.

An important ingredient for probing the link between SFR and interactions is the atomic and molecular gas content. 
Early work by \citet[][]{1994A&A...281..725C} reported an enhancement of the CO(1-0) luminosity (and SFR) 
in tidally perturbed objects, indicating that the mass of molecular gas is higher in interacting systems. 
Using a sample of 107 visually classified post-merger galaxies, 
\citet[][]{2018MNRAS.478.3447E} find that merged galaxies exhibit an 
atomic gas fraction enhancement compared with the 
control sample \citep[extended \emph{GALEX} Arecibo SDSS Survey;][]{2018MNRAS.476..875C} 
of the same stellar mass: they conclude that quenching is not a result 
of post-merger gas exhaustion \citep[see also][]{2018ApJ...868..132P}. 
\citet[][]{2019A&A...627A.107L} find no enhancement of the total H{\sc\,i}+H$_{2}$ gas mass fraction in major-merger pairs, 
relative to non-interacting comparison samples. 
\citet[][]{2016ApJ...825..128L} report a dependence of the molecular gas mass fraction on the merger classification stage. 
They postulate that interactions sweep the available atomic hydrogen from the galaxy outskirts into the central regions, 
where it is converted into H$_{2}$ and, eventually, into newly formed stars. 
In this process, the encounter geometry is an important factor for the gas inflow, 
as shown in the simulations of \citet[][]{2018MNRAS.479.3952B}. 
Analysis of interactions in the SIMBA cosmological simulation by \citet[][]{2019MNRAS.490.2139R} reveals that 
major mergers ($\le$ 4:1) induce SFR enhancements owing to an increase of the H$_{2}$ content at low masses, 
but when $M_{\ast}\ge 10^{10.5}M_{\odot}$ such an enhancement is attributed to a higher star formation (SF) efficiency associated with denser gas.

This work aims at understanding in more detail why galaxies exhibit extreme SFRs, 
whether they are interacting or not. Expanding the work by \citet[][]{2015MNRAS.454.1742K} and \citet[][]{2015ApJ...807L..16K}, 
we use their sample of $\sim 1500$ nearby galaxies drawn from the \emph{Spitzer} Survey of Stellar Structure in Galaxies \citep[S$^4$G;][]{2010PASP..122.1397S}. 
Without an anchoring of the analysis at cold gas masses, the desired comparison with theoretical predictions and simulations would remain incomplete. 
In this work we address this impediment using archival 21 cm H{\sc\,i} integrated data. 
An additional goal is to shed light on the physical properties of the starburst galaxies 
with an emphasis on the central stellar mass concentration and their gas content. 
%
\section{Sample and data}\label{data1}
%
Our parent sample is the S$^4$G, 
which is a magnitude- and diameter-limited survey that comprises 2352 galaxies with distances $\lesssim 40$ Mpc 
observed in the 3.6~$\mu$m and 4.5~$\mu$m bands with the Infrared Array Camera \citep[IRAC;][]{2004ApJS..154...10F} 
installed on board the \emph{Spitzer Space Telescope} \citep{2004ApJS..154....1W}. 

We use the catalogue of interacting and merging galaxies in  S$^4$G of \citet[][]{2014A&A...569A..91K}. 
Our sample comprises 1341 galaxies. Of these, 16 are currently merging (class A), 
39 appear highly distorted as a result of interaction with a companion (class B), and 84 present minor ongoing interaction (class C). 
The remaining 1202 galaxies constitute our control sample (CS) of non-interacting galaxies. 
We have excluded 138 galaxies from the \citet[][]{2014A&A...569A..91K} catalogue sample that belong to class "0", 
namely systems that have a close companion, but show no signs of interaction; 
this companion is within a radius of five times the diameter of the sample galaxy, 
has a recession velocity within $\pm 200$ km s$^{-1}$, and is not more than 3 mag fainter.
%
%
\begin{figure}
\centering   
\begin{tabular}{c c c}
   \includegraphics[width=0.5\textwidth]{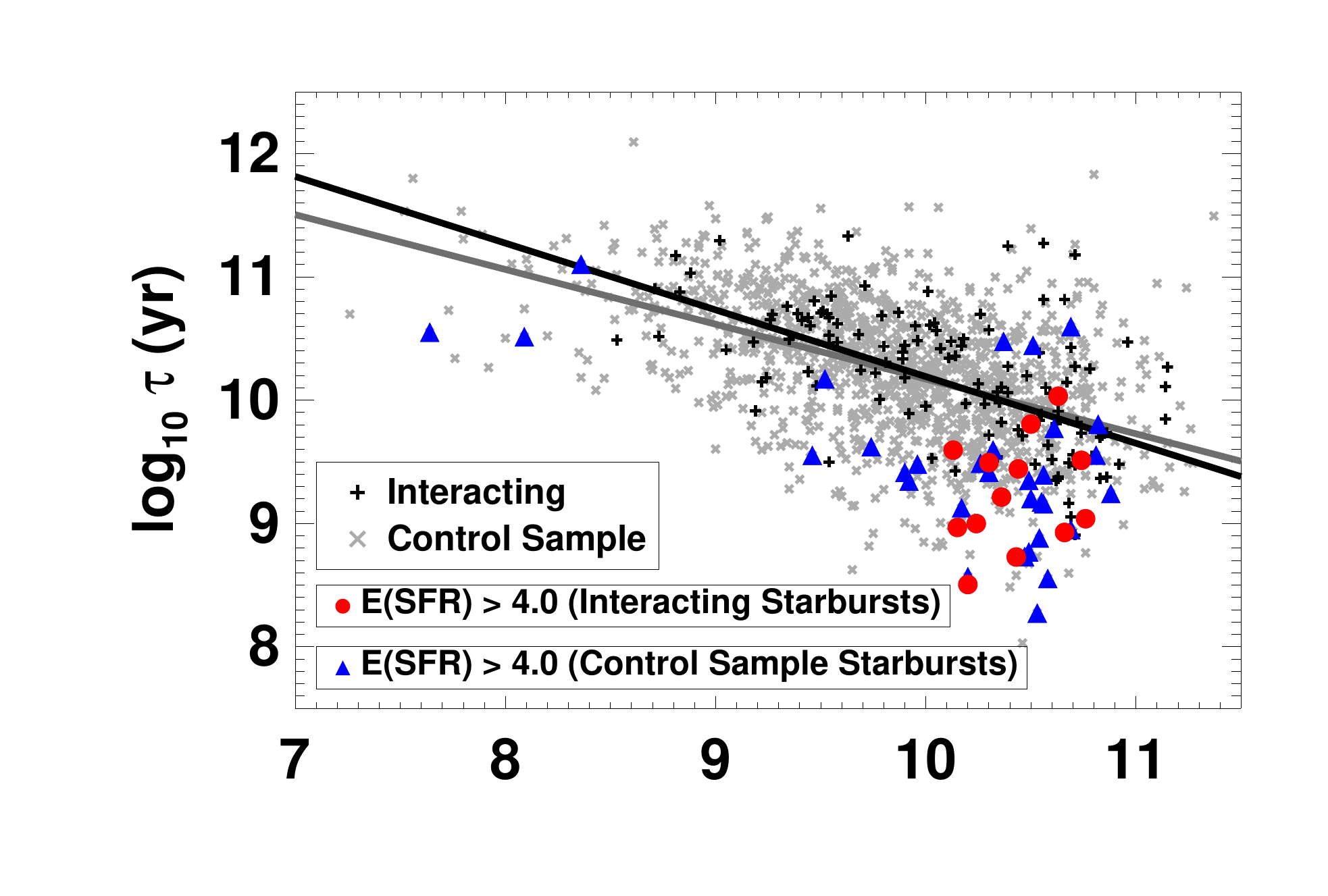}\\[-8ex]
   \includegraphics[width=0.5\textwidth]{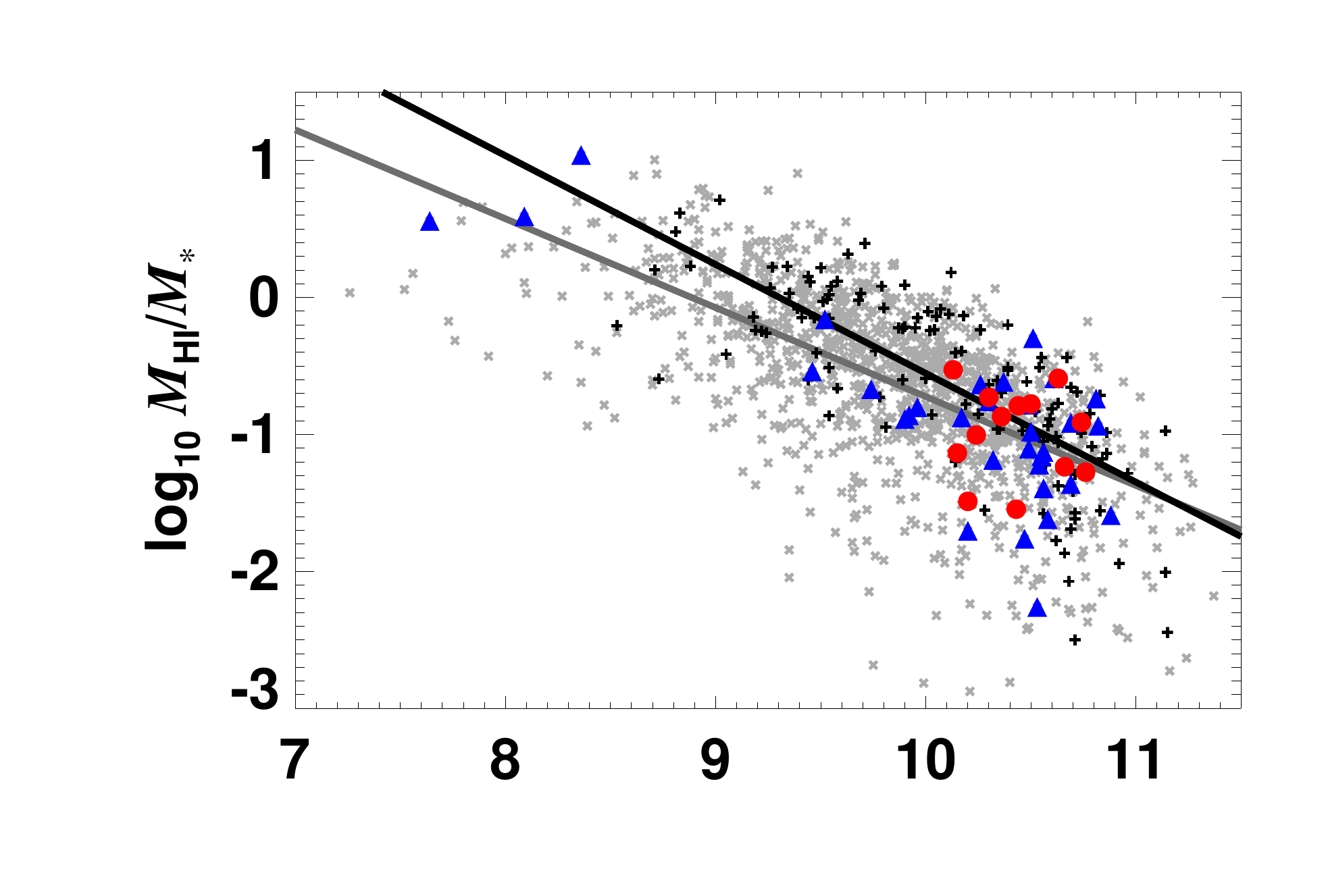}\\[-8ex]
   \includegraphics[width=0.5\textwidth]{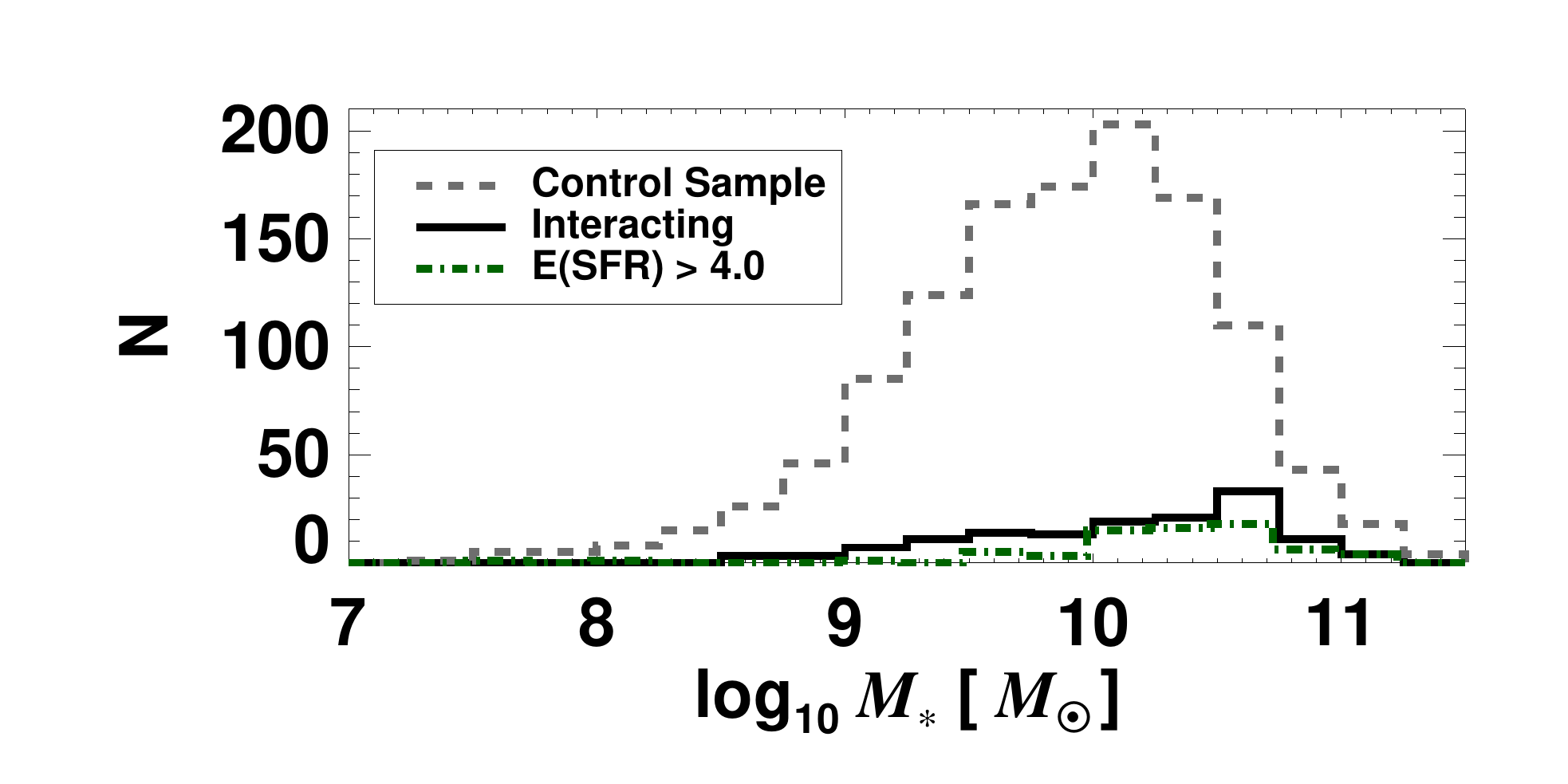}
\end{tabular}
\caption{
Gas depletion time (\emph{upper panel}) and gas fraction (\emph{middle panel}) as a function of total stellar mass. 
Non-interacting and interacting galaxies, according to \citet[][]{2014A&A...569A..91K}, 
are shown in grey and black, respectively. 
Non-interacting and interacting starbursts are shown in blue and red, respectively (see legend). 
The lines correspond to the linear fits to the galaxy points. 
In the \emph{lower panel} we represent  a histogram of galaxies vs. $M_{\ast}$ (bins of 0.25 dex). 
}
\label{fig_depletion_gas_frac_Mstar}
\end{figure}

The total SFRs used in this work are taken from \citet[][]{2015ApJS..219....5Q}, who calculated these values from the global 
\emph{IRAS} photometry at 60 $\mu$m and 100 $\mu$m, following \citet[][]{2000A&A...354..836L}. 
Total stellar masses ($M_{\ast}$) are taken from \citet[][]{2015ApJS..219....3M} and are used to calculate 
specific SFRs (sSFR=SFR/$M_{\ast}$). 
The same SFR values were applied by \citet[][]{2015MNRAS.454.1742K}, 
who used $M_{\ast}$ estimates from contaminant-free mass maps \citep[][]{2015ApJS..219....5Q} derived from 3.6 and 4.5 $\mu$m imaging. 
In this work we opt to consistently use the 3.6 $\mu$m passband for stellar mass inferences, including central concentration. 
Besides taking into account H{\sc\,i} information, a further improvement with respect to the work by \citet[][]{2015MNRAS.454.1742K} 
is our use of the more refined morphological classifications by \citet[][]{2015ApJS..217...32B}, 
instead of those from HyperLEDA \citep[][]{2003A&A...412...45P}. 
Atomic gas masses (in $M_{\odot}$) are estimated as \citep[e.g.][]{1988gera.book..522G,2018MNRAS.474.5372E,2019A&A...631A..94D} 
\begin{equation}\label{gasfrac}
M_{\rm HI}=2.356 \cdot 10^5 \cdot D^2 \cdot 10^{0.4 \cdot (17.4-m21c)},
\end{equation}
where $m21c$ is the corrected 21 cm line flux in magnitude from HyperLEDA\footnote{
We acknowledge the usage of the database http://leda.univ-lyon1.fr}
(available for 98$\%$ of the galaxies in our sample) 
and $D$ is the distance to the galaxy (in Megaparsec) adopted by \citet[][]{2015ApJS..219....3M}. 
From this, gas depletion times (in yr) are calculated as \citep[e.g.][]{2009ApJ...698.1437K}
\begin{equation}\label{gasfrac}
\tau=\dfrac{2.3 \cdot M_{\rm HI}}{0.6 \cdot \rm SFR},
\end{equation}
where the factor $2.3$ is applied to correct for helium and molecular gas content \citep[][]{2006ApJS..165..307M}, 
and the factor 0.6 accounts for the fraction of formed stars which is recycled to the interstellar medium \citep[][]{2008A&A...484..703J}.

We also use the inner gradient of the stellar component of the rotation curve \citep[$d_{\rm R}v_{\ast}(0)$, from][]{2016A&A...587A.160D} 
as a proxy of the central stellar mass concentration \citep[e.g.][]{2016MNRAS.458.1199E,2016A&A...596A..84D,2019A&A...625A.146D}. 
Specifically, $d_{\rm R}v_{\ast}(0)$ is obtained from a polynomial fit to the inner part of disc+bulge component of the circular velocity, 
following \citet[][]{2013MNRAS.433L..30L}, and taking the linear term as an estimate of the inner slope 
(80 of the 139 interacting and 768 of the 1202 non-interacting galaxies in our sample have reliable measurements of $d_{\rm R}v_{\ast}(0)$).

For every galaxy in our sample, we calculate the enhancement (E) in the global SFR, specific SFR, H{\sc\,i} gas mass, 
gas fraction, and gas depletion time by dividing these parameters by the median values for a control sample: 
$E$(SFR)=SFR/<SFR>$_{\rm cs}$, $E$(sSFR)=sSFR/<sSFR>$_{\rm cs}$, $E$($M_{\rm HI}$)=$M_{\rm HI}$/<$M_{\rm HI}$>$_{\rm cs}$, 
$E$($M_{\rm HI}/M_{\ast}$)=$[M_{\rm HI}/M_{\ast}]$/<$M_{\rm HI}/M_{\ast}$>$_{\rm cs}$, and $E$($\tau$)=$\tau$/<$\tau$>$_{\rm cs}$. 
A separate control sample is constructed for each individual galaxy, 
which encompasses all the non-interacting galaxies with similar values of $T$-type ($\pm 1$) and total stellar mass ($\pm$ 0.2 dex), 
following the criteria of \citet[][]{2015MNRAS.454.1742K}. 
Whenever the control sample of a certain galaxy has less than 5 objects (47 cases), this galaxy is not included in our statistics. 
The median number of galaxies in the control samples is 46. 
In 17 cases (16 from CS, 2 of class A) with uncertain $T$-type, we selected the control sample based on $M_{\ast}$ exclusively. 

Starbursts are defined as those galaxies having $E$(SFR)$>4$. 
Different SFR enhancement cut-offs (also on the sSFR) were tested by \citet[][their Fig.~1]{2015ApJ...807L..16K}, 
showing consistency in their result that the fraction of interacting galaxies is enhanced in starbursts. 
We checked and confirmed that the presented observational trends are 
qualitatively the same when starbursts are defined as $E$(SFR)$ > 4-5$ or $E$(sSFR)$ > 4-5$, fully in line with 
what \citet[][]{2015ApJ...807L..16K} reported.
%
\section{Results}\label{results}
%
%
\begin{figure}
\centering
\begin{tabular}{c c c}
   \includegraphics[width=0.5\textwidth]{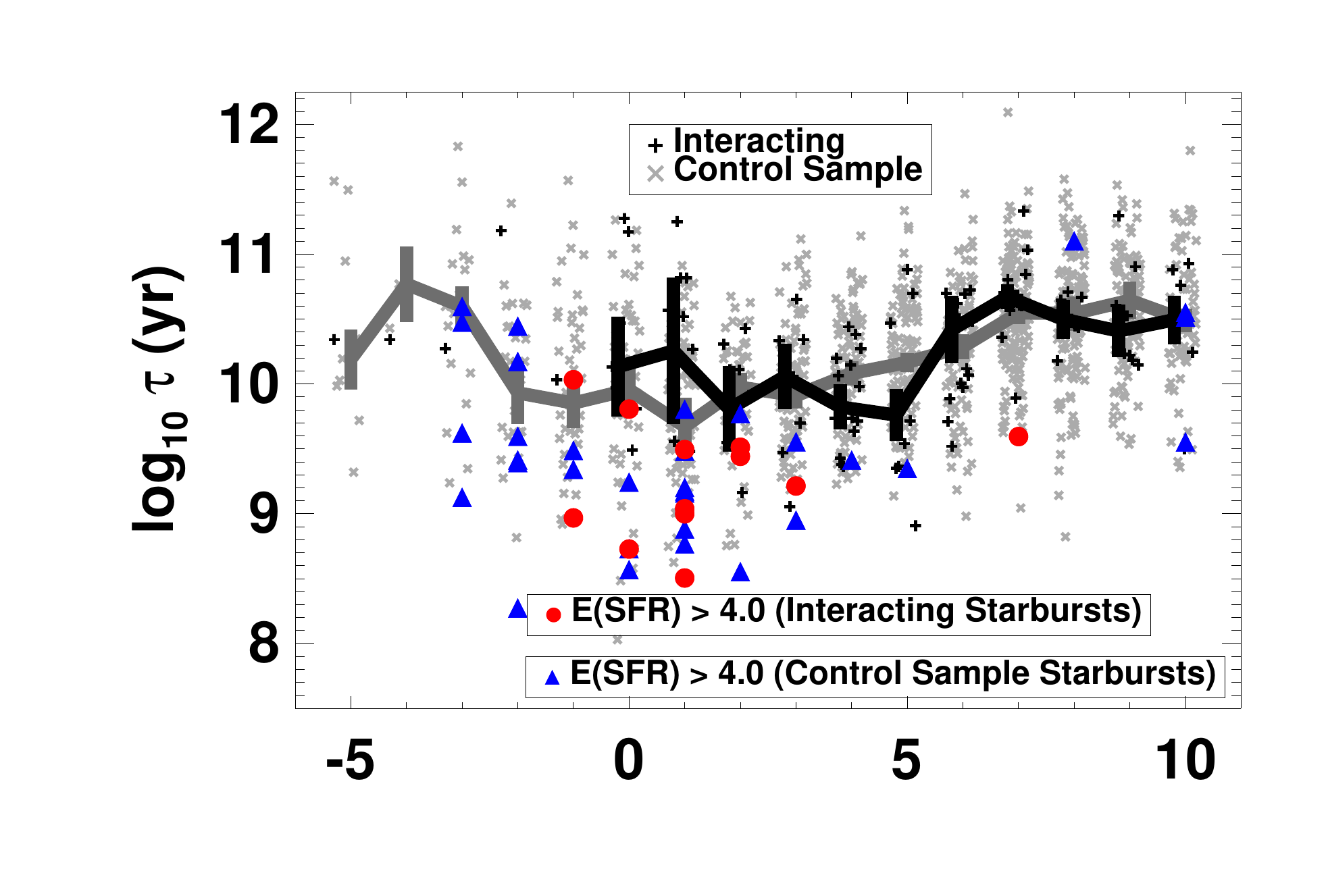}\\[-8ex]
   \includegraphics[width=0.5\textwidth]{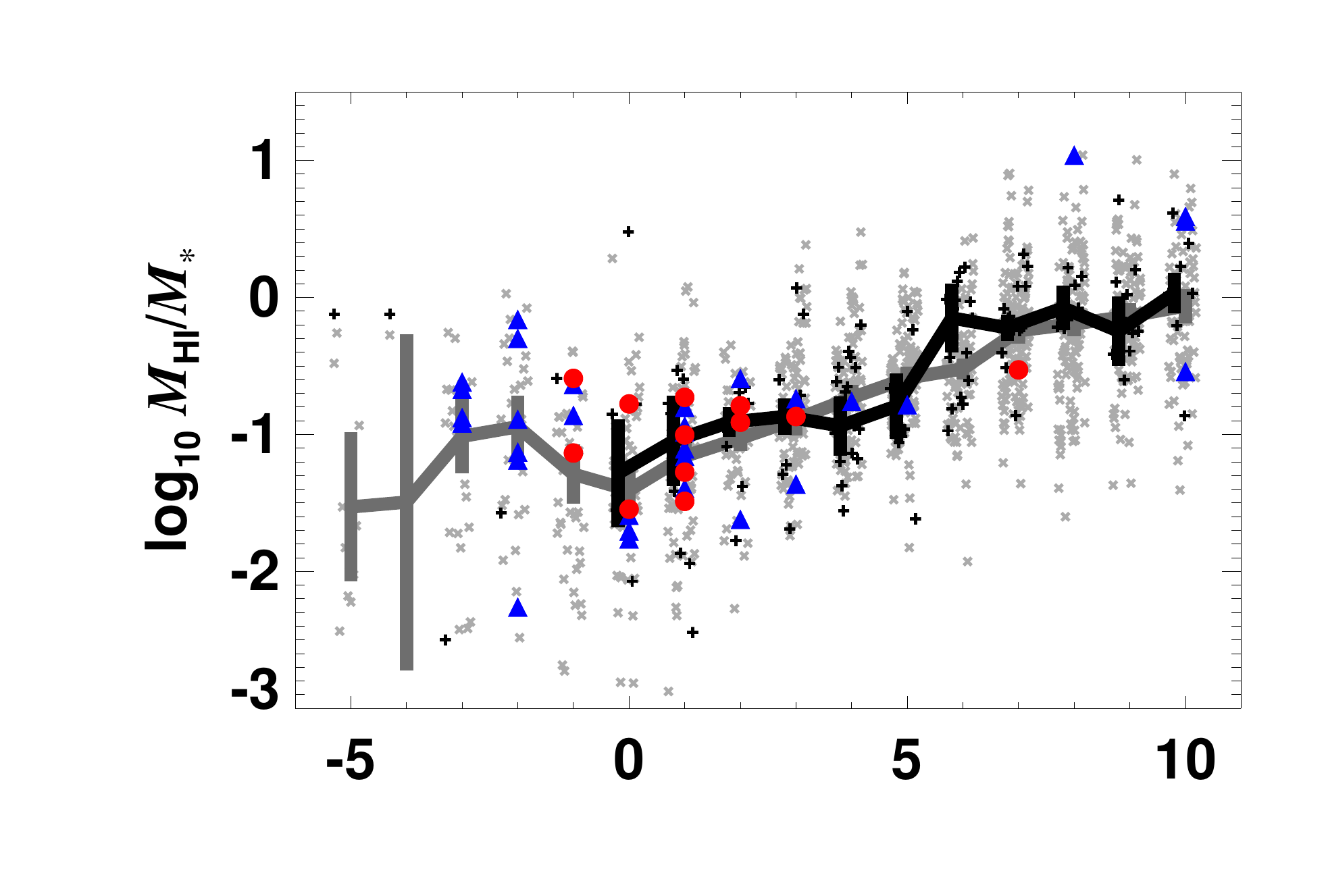}\\[-8ex]
   \includegraphics[width=0.5\textwidth]{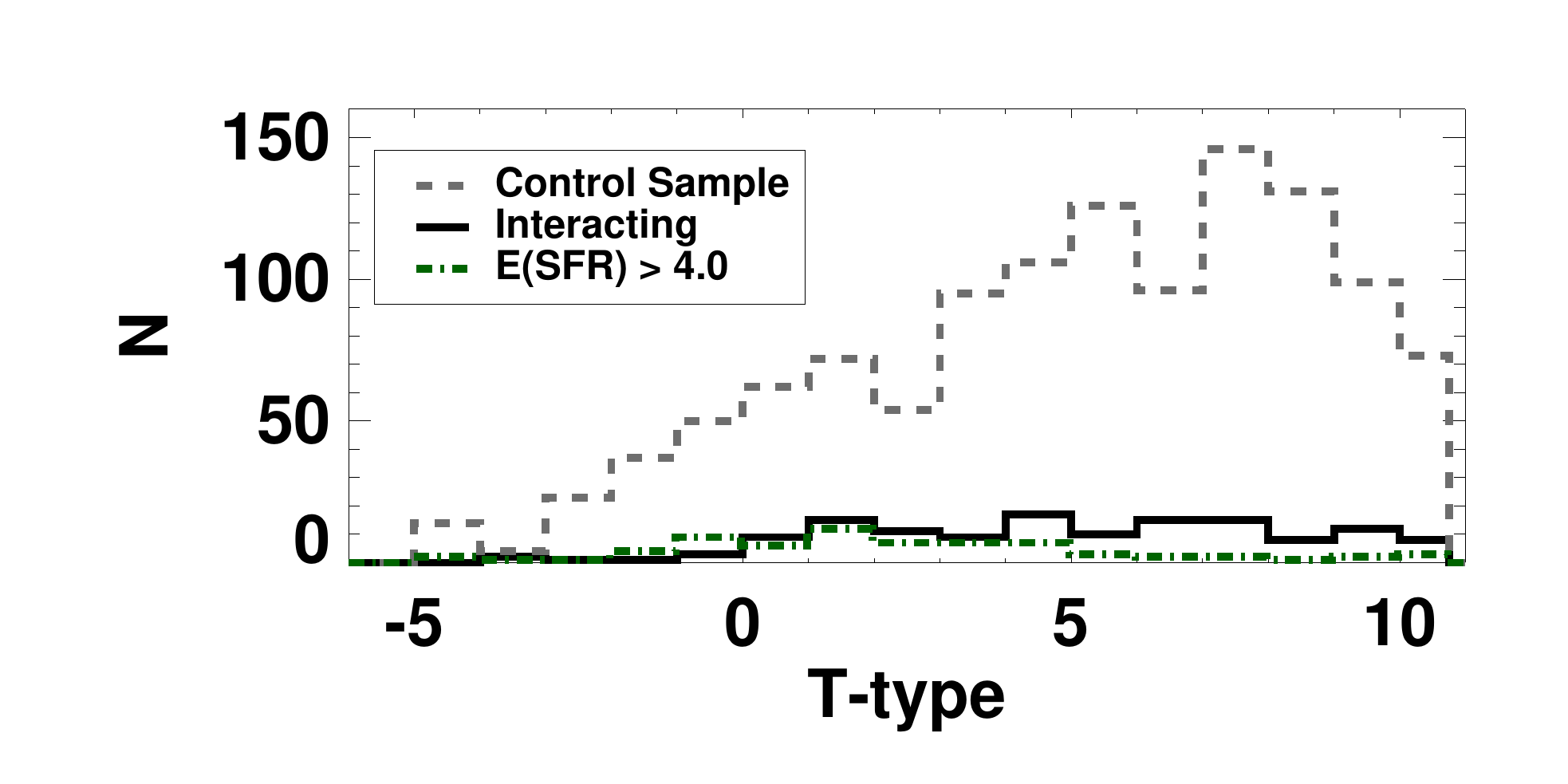}
\end{tabular}
\caption{
As in Fig.~\ref{fig_depletion_gas_frac_Mstar} but as a function of the revised Hubble stage. 
The running median (error bars obtained via bootstrap re-samplings) is shown for interacting (black) and non-interacting (grey) 
galaxies separately. In the upper and central panels we added small random offsets ($\lesssim 0.3$) to the 
$T$ values in the $x-$axis (integers) to avoid point overlapping.
}
\label{fig_depletion_gas_frac_T_Type}
\end{figure}
Depletion times and gas fractions decrease with increasing total stellar mass (Fig.~\ref{fig_depletion_gas_frac_Mstar}). 
When studied in the Hubble sequence, $\tau$ and $M_{\rm HI}/M_{\ast}$ are larger for larger $T$ among the spirals 
(Fig.~\ref{fig_depletion_gas_frac_T_Type}). Curiously enough, $\tau$ seems to be larger than average among lenticulars, 
although the sampling of S0s is not optimal because of the bias of the S$^4$G towards late-type, gas-rich systems.

For a given $T$ or $M_{\ast}$ bin, the average $\tau$ and $M_{\rm HI}/M_{\ast}$ are 
not significantly different for interacting and non-interacting galaxies 
(black and grey symbols and lines, respectively), 
as seen in both Fig.~\ref{fig_depletion_gas_frac_Mstar} and Fig.~\ref{fig_depletion_gas_frac_T_Type}. 
To confirm this result, we performed a two-sample Kolmogorov-Smirnov (K-S) test 
\citep[IDL implementation \emph{kstwo.pro} written by W. Landsman, following][]{1986nras.book.....P}, 
finding that the cumulative distribution function of $\tau$ and $M_{\rm HI}/M_{\ast}$ 
is similar for interacting and non-interacting galaxies, as indicated by the large ($> 0.01$) $p-$values ($0.27$ and $0.46$, respectively).  

The distribution of $M_{\ast}$ for non-interacting galaxies 
peaks at $10^{10-10.25}M_{\odot}$ (lower panel of Fig.~\ref{fig_depletion_gas_frac_Mstar}), 
while for interacting systems it peaks at slightly larger masses $M_{\ast}\approx 10^{10.5-10.75}M_{\odot}$. 
Interacting galaxies spread evenly across the Hubble sequence (lower panel of Fig.~\ref{fig_depletion_gas_frac_T_Type}), 
but only 5 S0s in our sample are interacting. 
K-S tests indicate that the two arrays of $M_{\ast}$ and $T$ values for interacting and CS galaxies 
are drawn from different distributions ($p-$values of $3.8 \cdot 10^{-5}$ and $3.7 \cdot 10^{-3}$, respectively).

\begin{figure*}
\centering   
\begin{tabular}{c c}
   \includegraphics[width=0.99\textwidth]{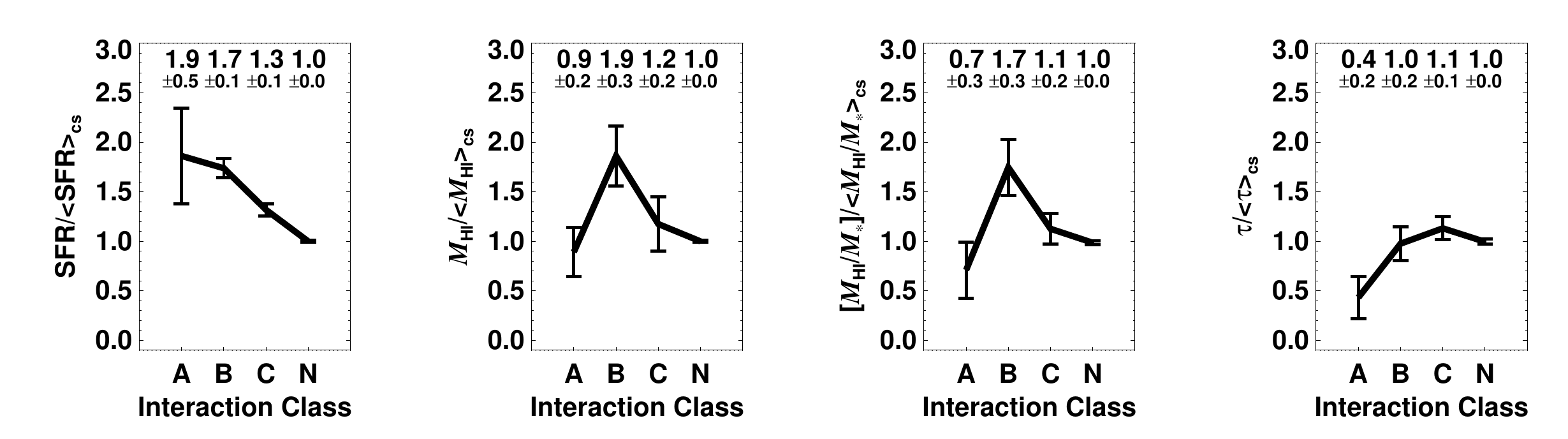}\\
\end{tabular}
\caption{
Interaction class (A = merging, B = highly distorted due to interaction, C = minor interaction, and N = control sample) 
vs.\ SFR, gas mass, gas fraction, and depletion time, all normalised to CS median values (from left to right). 
We show the running median and confidence limit based on bootstrap re-samplings (values are shown at the top). 
}
\label{fig_gas_interact}
\end{figure*}

\begin{figure*}
\centering   
   \includegraphics[width=0.99\textwidth]{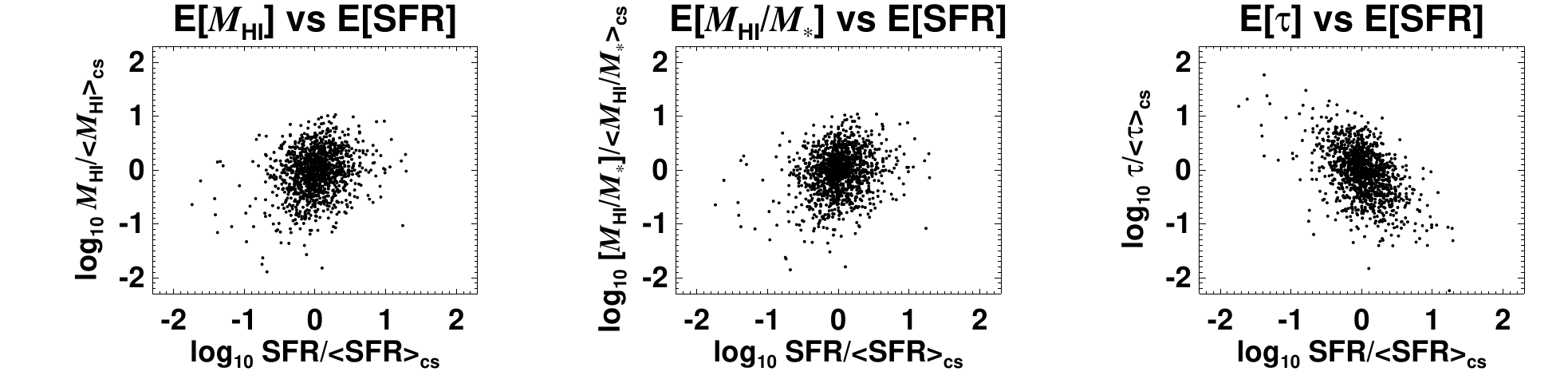}
\caption{
Enhancement of H{\sc\,i} gas masses (left), gas fractions (centre), 
and gas depletion times (right) as a function of enhancement on SFRs. 
}
\label{boost_comparison}
\end{figure*}

Fig.~\ref{fig_gas_interact} shows the enhancements on the SFR, H{\sc\,i} gas mass and gas fraction (normalised by $M_{\ast}$), 
and gas depletion timescale for multiple interaction classes. 
Median factors (and uncertainties obtained via bootstrap re-samplings) are indicated at the top of the panels. 
The median enhancement in SFR (E(SFR)) increases with increasing degree of interaction, 
where class A presents a median factor of $1.9\pm0.5$ higher SFR than the median for the CS galaxies 
\citep[in agreement with][]{2015MNRAS.454.1742K}. 
Likewise, highly interacting systems are characterised by a median E($\tau$) of $0.4\pm 0.2$. 
Regarding atomic gas masses, class B has median E($M_{\rm HI}$) and E($M_{\rm HI}/M_{\ast}$) of $1.9\pm0.3$ and $1.7\pm 0.3$, respectively, 
but classes A and C do not show significant differences as compared to the CS. 
Extreme enhancements can be found in galaxies with different degrees of interaction or in complete isolation. 
Finally, enhancements on $M_{\rm HI}$, $M_{\rm HI}/M_{\ast}$, and $\tau$ are compared to that of the SFR in Fig.~\ref{boost_comparison}. The quantities 
E($M_{\rm HI}$) and E($M_{\rm HI}/M_{\ast}$) hardly show any correlation with E(SFR), as indicated by Spearman's correlation coefficients 
(significances) $\rho=0.19$ ($p=3.6 \cdot 10^{-12}$) and $\rho=0.18$ ($p=6 \cdot 10^{-11}$), respectively, 
while a more robust correlation between E(SFR) and E($\tau$) is found ($\rho=-0.45$, $p=0$, naturally expected from the definition of $\tau$).
%
%
%
%

There are 45 starburst galaxies in our sample, i.e. fulfilling $E$(SFR)>4. 
These are typically massive systems with $M_{\ast}\gtrsim 10^{10}M_{\sun}$ (Fig.~\ref{fig_depletion_gas_frac_Mstar}) 
and Hubble stages $T\lesssim 5$ (Fig.~\ref{fig_depletion_gas_frac_T_Type}) ($80\%$ of the cases). 
For a given $M_{\ast}$ or $T$ bin, starbursts present lower gas depletion timescales than the median. 
However, their gas fractions are not larger than the median. This means that the lower $\tau$ is only determined by the enhanced SFR. 
There are three massive systems in our sample (NGC$\,$2894, IC$\,$2461, and UGC$\,$07522) 
that present high depletion timescales even though they are starbursts. All three of these massive systems are S0$^{-}$ and S0$^{0}$ galaxies.

\begin{figure}
\centering   
   \includegraphics[width=0.5\textwidth]{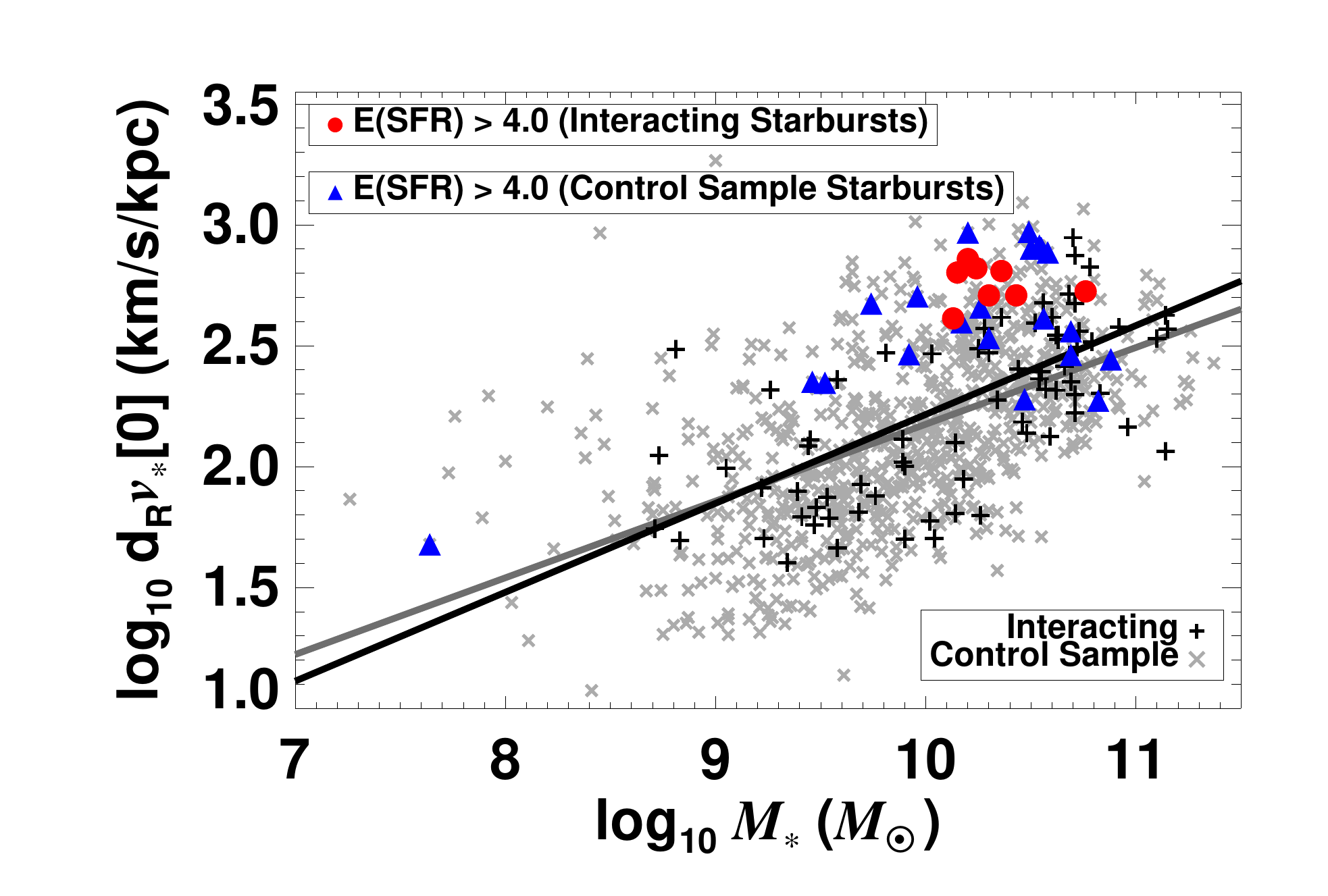}
\caption{
Inner slope of the stellar component of the rotation curve (direct measurement of the central stellar mass concentration) 
vs. total stellar mass for interacting (black) and non-interacting (grey) galaxies, 
while highlighting the starbursts (blue and red). The lines correspond to the linear fit to the data clouds. 
}
\label{fig_inner}
\end{figure}

Starbursts are characterised by higher central concentrations of stars (Fig.~\ref{fig_inner}): 
for a given $M_{\ast}$ bin, galaxies with $E$(SFR)$>4$ show higher values of $d_{\rm R}v_{\ast}(0)$. 
Interacting and non-interacting galaxies are not significantly different in $d_{\rm R}v_{\ast}(0)$ for a certain $M_{\ast}$. 
However, as interacting galaxies are in general slightly more massive than their CS counterparts, 
their global $d_{\rm R}v_{\ast}(0)$ distribution peaks for slightly higher values, 
and thus the significance level of the K-S statistic between the subsamples of interacting and CS galaxies is fairly low 
($p$-value of 0.0036). 
%
%
\section{Discussion}\label{discussion}
%
%
Using refined morphological types \citep[][]{2015ApJS..217...32B} for the control sample definition, 
we confirm the result by \citet[][]{2015MNRAS.454.1742K} that the enhancement in SFR for highly interacting galaxies is 
a factor of $\sim 2$ higher than the average for CS galaxies. 
Our results are consistent with the expectations from the simulations by \citet[][]{2007A&A...468...61D,2008A&A...492...31D}, 
who showed a similar enhancement of (circumnuclear) SF due to galaxy collisions, which drive the formation of non-axisymmetries that trigger inward gas flows \citep[][]{1991ApJ...370L..65B,1996ApJ...464..641M}.

In this work we take a step forward by adding the cold gas masses to the analysis for the different interacting classes. 
Studying all subclasses of interacting galaxies separately is undoubtedly insightful, 
yet uncertainties associated with the beam size for the far-infrared (\emph{IRAS}) and 21 cm data (data from different telescopes are gathered in LEDA) 
may affect the measurements, and thus care must be taken to tone down conclusions based on interacting class. 
In certain cases (class A) strongly interacting galaxies may be so close that they could have been observed together, 
resulting in artificially increased values of $M_{\rm HI}$, SFR, or $\tau$. We argue that such effects do not strongly affect 
our statistics when all interacting classes are grouped.

\citet[][]{2009ApJ...698.1437K} showed that H{\sc\,i} total and relative mass are not strongly dependent on 
the absence or presence of a close companion, but do depend strongly on $T$-type. 
This is confirmed in this work with a four times larger sample (Fig.~\ref{fig_depletion_gas_frac_T_Type}). 
Numerical models by \citet[][]{2007A&A...468...61D} showed that the gas content in the merging event 
is not the main parameter governing the SF efficiency. 
Observational work by \citet[][]{2018MNRAS.476.2591V} finds that gas fractions and depletion times 
in galaxy pairs are consistent with those of non-mergers whose SFRs are similarly raised. 
In addition, recent work by \citet[][]{2018MNRAS.478.3447E} shows that there is no correlation between the H{\sc\,i} gas fraction 
enhancement of a galaxy and its SFR enhancement (see their Fig.~7). We assessed this in our sample 
and showed that, indeed, E(SFR) is barely correlated with E($M_{\rm HI}$) and E($M_{\rm HI}/M_{\ast}$) (Fig.~\ref{boost_comparison}). 

Interestingly, we find that in highly distorted galaxies (class B) the H{\sc\,i} mass is significantly enhanced 
(almost a factor of 2 higher than the CS, on average). 
This is not the case in class A, where the merger-driven enhanced SFR (plausibly prolonged for several billion years) 
might have started to quench the galaxy, given the low depletion timescales. 
On the other hand, the cold gas could have been heated or removed via feedback from active 
galactic nuclei \citep[e.g.][and references therein]{2018NatAs...2..198H,2019MNRAS.482.5694E}. 
Also, \citet[][]{2007A&A...468...61D} found that strong tidal interactions can remove a large amount of gas from 
the galaxy discs that is not fully re-acquired in the last merging phases. 
Finally, we note that the correlation between E(SFR) and E($\tau$) is not very tight (Fig.~\ref{boost_comparison}), 
which emphasises how controversial the definition of starburst can be \citep[see][]{2009ApJ...698.1437K}.

In class A galaxies (mergers), 
we find the median $\tau$ to be a factor of $0.4 \pm 0.2$ that of the CS. 
This is consistent with the findings by \citet[][]{2012ApJ...758...73S}, 
who showed that among the gas-rich disc-dominated population galaxies 
those undergoing mergers or presenting disturbed morphologies are characterised by short depletion times, 
roughly a factor of two shorter when compared to the CS. 
Saintonge et al. also note that not all mergers or interactions are necessarily associated with episodes of efficient SF, 
which is also the case in our work; we also find low values of E($\tau$) among all interacting classes.

Our analysis reveals that starbursts are characterised by a higher stellar mass central concentration when compared to a control sample, 
as measured from 3.6 $\mu$m imaging. 
This is most likely related to the fact that these galaxies have mainly harboured SF in the circumnuclear regions. 
This may imply that the SF enhancement is not associated with a single instantaneous central SF burst, 
but instead with either continuous SF over a period of $10^{8-9}$ yr \citep[e.g.][]{2009ApJ...698.1437K}, 
probably resulting from large-scale gas inflow driven by stellar non-axisymmetries, 
or a series of bursts over a significant timescale \citep[e.g.][]{2006MNRAS.371.1087A}. 
In addition, gas depletion times are longer in bulge-dominated galaxies according to \citet[][]{2012ApJ...758...73S}. 
However, we checked that the $M_{\ast}$-$\tau$ relation is not segregated as a function of $d_{\rm R}v_{\ast}(0)$ 
\citep[the bulge mass is larger for larger $d_{\rm R}v_{\ast}(0)$;][]{2016PhDT.......168D}. 
From SDSS-IV MaNGA data, \citet[][]{2019ApJ...881..119P} suggest 
that interaction-triggered SF is not necessarily limited to the circumnuclear regions, but the enhancement is centrally peaked. 
Recent work by \citet[][]{2020MNRAS.tmp...36E} using MaNGA and ALMA data for 12 galaxies 
shows that an elevated SF efficiency is the fundamental driver for central starbursts, which have contributions from both interactions and secular mechanisms. 
In our work, high central stellar mass concentrations are detected for most starbursts regardless of whether they are interacting or not. 
%
%
%
%
%
\section{Summary and conclusions}\label{conclusion}
%
%
We used a sample of 1341 S$^4$G galaxies from the catalogue on interacting and merging galaxies produced by \citet[][]{2014A&A...569A..91K} 
to analyse the effect of galaxy-galaxy interactions on SFRs, 
atomic gas masses ($M_{\rm HI}$) and fractions ($M_{\rm HI}/M_{\ast}$), and depletions times ($\tau$), using archival 21 cm H{\sc\,i} 
data from HyperLEDA and \emph{IRAS} far-infrared fluxes \citep[][]{2015ApJS..219....5Q}. 
We also studied the global physical properties of starburst galaxies, 
such as total stellar masses \citep[$M_{\ast}$, from][]{2015ApJS..219....3M} and central concentration 
\citep[traced from the inner slope of the stellar component of the rotation curves, 
$d_{\rm R}v_{\ast}(0)$, using 3.6~$\mu$m imaging, from][]{2016A&A...587A.160D}. 
Starbursts are defined as galaxies that have a factor $>4$ enhanced SFR relative to a control sample, 
which comprises non-interacting systems with $\pm 0.2$ dex in $M_{\ast}$ and $\pm 1$ in $T$ \citep[types from][]{2015ApJS..217...32B}. 
The main results of this paper are the following:
\begin{itemize}
\item We confirm that the distribution of gas fraction $M_{\rm HI}/M_{\ast}$ and depletion time $\tau$ 
is similar in interacting and control sample galaxies, but depends on $M_{\ast}$ and $T$ \citep[e.g.][]{2009ApJ...698.1437K}. 
\item In merging galaxies, the median enhancement in $\tau$ and SFR is a factor $0.4 \pm 0.2$ and $1.9 \pm 0.5$ 
that of the average for control sample galaxies, 
respectively \citep[in line with earlier work by e.g.][]{2012ApJ...758...73S,2015MNRAS.454.1742K}.
\item Highly distorted interacting galaxies have a factor of $1.9 \pm 0.3$ enhanced H{\sc\,i} mass. 
This is not the case in merging systems, where the (interaction-triggered) 
higher star formation efficiency might have started to quench the galaxies.
\item Starbursts are typically early-type massive galaxies ($M_{\ast}\gtrsim 10^{10}M_{\sun}$, $T \lesssim 5$) 
and can be either interacting or non-interacting. 
\item For a given $M_{\ast}$ bin, 
starbursts present lower gas depletion time $\tau$ than the average, yet they have similar $M_{\rm HI}/M_{\ast}$. 
An enhancement of the SFR does not imply an enhancement of the relative and absolute amount of gas, or vice versa.
\item For a given $M_{\ast}$ bin, starbursts are characterised by higher central stellar concentrations. 
This points to these systems having undergone continuous circumnuclear star formation over a period of $10^{8}$ to $10^9$ yrs, 
nourished by gas inflow that is driven by both interactions and non-axisymmetries.
\end{itemize}
%
%
This work highlights how the study of large comprehensive samples of nearby galaxies with accurately determined 
physical properties allows us to shed light on the boost of SF that certain galaxies experience and, in particular, 
the connection of this boost with galaxy-galaxy interactions. 
Specifically, it provides observational evidence that starbursts harbour long-lasting SF in circumnuclear regions, 
independent of whether they are interacting or not, and that galaxy mergers produce a moderate enhancement (by a factor of $\sim 2$) of the SF efficiency, 
in line with theoretical predictions that are based on numerical models.
%
%
\begin{acknowledgements}
We thank the anonymous referee for comments that improved this paper. 
We acknowledge financial support from the European Union's Horizon 2020 research and innovation programme under 
Marie Sk$\l$odowska-Curie grant agreement No 721463 to the SUNDIAL ITN network, 
from the State Research Agency (AEI) of the Spanish Ministry of Science, 
Innovation and Universities (MCIU) and the European Regional Development Fund (FEDER) under the grant with reference AYA2016-76219-P, 
and from the IAC project P/300724 which is financed by the Ministry of Science, Innovation and Universities, through the State Budget and by the 
Canary Islands Department of Economy, Knowledge and Employment, through the Regional Budget of the Autonomous Community. 
This project has received funding from the European Union’s Horizon 2020 research and innovation programme 
under the Marie Sk$\l$odowska-Curie grant agreement No 893673. 
JHK acknowledges financial support from the Fundaci\'on BBVA under its 2017 programme of assistance to scientific research groups, 
for the project "Using machine-learning techniques to drag galaxies from the noise in deep imaging". 
We thank Ute Lisenfeld for comments on the manuscript, and Miguel Querejeta for fruitful discussions. 
{\it Facilities}: \emph{Spitzer} (IRAC).
%
%
\end{acknowledgements}
\bibliographystyle{aa}
\bibliography{bibliography}
\clearpage
%
%
%
%
\end{document}